\documentclass{PoS}
\usepackage{epsfig}
\usepackage{cite}  
\newcommand{\be}{\begin{equation}}
\newcommand{\ee}{\end{equation}}
\newcommand{\bea}{\begin{eqnarray}}
\newcommand{\eea}{\end{eqnarray}}
\newcommand{\ba}{\begin{array}}
\newcommand{\ea}{\end{array}}

\newcommand{\fpi}{f_{\pi}}
\newcommand{\owt}{\omega_{\rm WT}}

\newcommand{\cowt}{\cot\omega_{\rm WT}}

\newcommand{\pih}{\pi/2}

\newcommand{\pref}[1]{(\ref{#1})}

\PoS{PoS(LAT2005)046}

\title{Determining the low energy parameters of Wilson Chiral Perturbation Theory }

 \ShortTitle{Determining the low energy parameters of WChPT }

\author{Sinya Aoki and \speaker{Oliver B{\"a}r} \\
Graduate School of Pure and Applied Sciences\\
University of Tsukuba\\
Tsukuba, Ibaraki  305-8571\\
Japan\\
E-mail: \email{saoki@het.ph.tsukuba.ac.jp},
\email{obaer@het.ph.tsukuba.ac.jp}}

\abstract{
We report preliminary results of a Wilson Chiral Perturbation Theory (WChPT) analysis of twisted mass lattice QCD data. The quenched data, previously published by two different groups, was generated with two definitions for the critical quark mass and shows a strong non-linear quark mass dependence for small quark masses for the pion mass definition (``bending phenomenon''). We find that WChPT describes this characteristic curvature fairly well. Fits to the data provide estimates for combinations of low-energy parameters, even though the errors are sizable.
}

\FullConference{XXIIIrd International Symposium on Lattice Field Theory\\
		 25-30 July 2005\\
		 Trinity College, Dublin, Ireland}
\begin{document}
%

%
\section{Introduction}
%
Twisted mass Lattice QCD (tmLQCD) \cite{Frezzotti:2000nk,Frezzotti:2001ea} is a promising formulation to approach the chiral limit of QCD with Wilson fermions. It solves the problem of exceptional configurations  and makes numerical simulations with small quark masses feasible. Recent quenched studies \cite{Bietenholz:2004wv,Abdel-Rehim:2005gz,Jansen:2005gf,Jansen:2005kk}
 reached
 values for $m_{\pi}/m_{\rho}$ as small as 0.3 without running into problems due to exceptional configurations. Besides this numerical advantage  tmLQCD has the property of automatic O($a$) improvement at maximal twist \cite{Frezzotti:2003ni}. More recent results in tmLQCD have been reviewed at this conference \cite{ShindlerLat2005}.  

Some issues, however, remain to be fully understood. In order to be at maximal twist the bare untwisted mass $m_{0}$ needs to be tuned properly.  It has been argued in Ref.\ \cite{Aoki:2004ta} that automatic O($a$) improvement is lost if one tunes to the critical quark mass where the  pion mass vanishes, unless the twisted quark mass satisfies the bound $\mu > a^{2}\Lambda_{\rm QCD}^{3}$. On the other hand, automatic O($a$) improvement is expected to hold if $m_{0}$ is tuned such that the PCAC quark mass vanishes. However, the results  in Ref.\ \cite{Aoki:2004ta} are controversial.
It is claimed \cite{Jansen:2005kk,Frezzotti:2005gi} that both definitions for the critical quark mass guarantee automatic O($a$) improvement, but the remaining O($a^{2}$) effects differ significantly. In particular, the pion mass definition is afflicted with cut-off artifacts of O($a^{2}/m_{\pi}^{2})$ which are enhanced for small pion masses \cite{Frezzotti:2005gi}. These enhanced lattices artifacts are shown to be reduced for the PCAC quark mass definition. 

Closely related is the so-called 'bending phenomenon', observed in quenched simulations \cite{Bietenholz:2004wv,Abdel-Rehim:2005gz,Jansen:2005gf,Jansen:2005kk}. Various mesonic quantities like the pion mass, the pion decay constant and the vector meson mass show an unexpected non-linear quark mass dependence for small quark masses if the pion mass definition for the critical quark mass is used.
This curvature is absent when the PCAC quark mass definition is employed.

Here we report  preliminary results of a Wilson Chiral Perturbation Theory (WChPT) analysis of the lattice data in Refs.\cite{Bietenholz:2004wv,Abdel-Rehim:2005gz,Jansen:2005gf,Jansen:2005kk}.  Our motivation for this analysis is two-fold. Firstly, we would like to check whether WChPT is able to describe the lattice data and the peculiar bending, as it should if it is the correct low-energy theory of tmLQCD. Previous analyses with WChPT \cite{Farchioni:2003nf,Aoki:2003yv,Namekawa:2004bi}, applied to data obtained with untwisted Wilson fermions, 
are inconclusive and do not provide sufficient evidence for the validity of WChPT. Secondly, provided WChPT can be applied, a fit to the data should give us some combinations of low-energy coefficients of WChPT. Particularly interesting is the coefficient $c_{2} $  which plays a crucial role for  the phase structure of the lattice theory \cite{Sharpe:1998xm,Munster:2004am,Sharpe:2004ps}. This coefficient can be determined by measuring the mass difference between the charged and neutral pion \cite{Scorzato:2004da}, however, this calculation is difficult in practice since it involves the computation of disconnected quark diagrams. A recent computation of the pion mass difference \cite{Jansen:2005cg} shows large statistical errors and alternative methods to get a handle on $c_{2}$ are clearly welcome.  
%
\section{WChPT for twisted mass LQCD}
%
We extended the  WChPT analysis in Ref. \cite{Aoki:2004ta} by including the terms of O($\mu a$) and O($a^{3}$) 
in the chiral Lagrangian. Keeping these terms is consistent in a power counting scheme where 
the renormalized masses $m$ and $\mu $ and the lattice spacing $a$  satisfy 
(proper powers of $\Lambda_{\rm QCD}$ are
, as usual, 
understood)
\be
\mu \approx a^{2} > \mu a \approx a^{3} > \mu^{2} \approx \mu a^{2}\approx a^{4}.
\ee
Even though the O($\mu a,a^{3})$ contributions are subleading, we found them to be essential to describe some features in the lattices data.

A twisted mass term breaks the flavor symmetry. The ground state $\Sigma_{0}$ of the theory is therefore no longer symmetric under flavor rotations but given by 
\bea
\Sigma_{0} &=& \exp (i \phi \tau_3),
\eea
where $\phi$ denotes the vacuum angle. Minimizing the potential energy we derive a {\em gap equation},
\bea\label{SquaredGapEQ}
\alpha^{2}\Big( t -\beta(1-2t^{2})\Big)^2 &=& \Big(\chi-t +2\beta \chi_{m}t +\gamma t^{2}\Big)^2(1-t^2),
\eea
which determines $t=\cos\phi$ as a function of $\mu$, $m$ and  $a$. The masses and the lattice spacing are included in the dimensionless parameters
\bea
\alpha&=&\frac{2B\mu}{2c_{2}a^{2}},\quad
\chi \,=\, \frac{ 2 B m + 2 W_{0}a(1 + \tilde{c}_{3}a^{2}) }{2c_2a^{2}},\quad
\chi_{m}\, =\, \frac{ 2 B m  }{2c_2a^{2}}, \quad
\beta\, = \, \frac{\tilde{c}_{2}a}{2B},\quad 
\gamma \,=\, \frac{3c_{3}a^{3}}{2c_{2}a^{2}},
\eea 
where $B,W_{0}$, $c_{i}$, $\tilde{c}_{i}$ are (combinations of) low-energy parameters in the chiral Lagrangian.
Expanding in terms of pion fields in the usual way one derives the pion mass, the pion decay constant and the twist angle $\owt$ stemming from the chiral Ward identities \cite{Frezzotti:2000nk}:
\bea
m_{\pi_{ \pm}}^2 &=&\frac{2B\mu}{\sqrt{1-t^2}}\frac{1 + \beta t}{1+\delta_{1}t}\left[1- \delta\ln\left(\frac{2B\mu}{\Lambda_{\chi}^{2}}\right)\right],\label{Mpi}\\
f_{\pi}& =& f (1+ \delta_{1}^{\prime} t + \delta_{\mu}\mu \sqrt{1-t^{2}})\sqrt{1-t^{2}},\\
\cowt & = & \frac{t}{\sqrt{1-t^{2}}},\label{Def:cotw}
\eea
where $\delta_{1},\delta^{\prime}_{1},\delta_{\mu}$ are additional combinations of low-energy coefficients  and $\delta$ multiplies the quenched chiral log contribution \cite{Bernard:1992mk,Sharpe:1992ft}. For $\fpi$ we also included the O($\mu^{2}$) correction which we dropped in the result for the pion mass.

The results \pref{Mpi} - \pref{Def:cotw} are valid for arbitrarily chosen twisted and untwisted quark masses. In order to make contact to numerical lattice data one has to properly match the untwisted quark mass according to the chosen definition for the critical quark mass. With the so-called parity conservation definition \cite{Farchioni:2004fs,Sharpe:2004ny} one tunes $m$ such that $\owt = \pi/2$, which implies $t=0$. In this case the expressions for the pion mass and decay constant reduce to what one expects from continuum ChPT. With the pion mass definition one sets the bare untwisted mass to the value where the pion becomes massless (for vanishing 
$\mu$). In this case the gap equation implies that $t$ depends on $\alpha$ and therefore on $\mu$.  Hence the presence of $t$ in  \pref{Mpi} - \pref{Def:cotw} causes deviations from the continuum ChPT expressions. 

 \begin{figure}[t]
 \vspace{-0.4cm}
 \begin{center}
       \scalebox{0.63}{\includegraphics{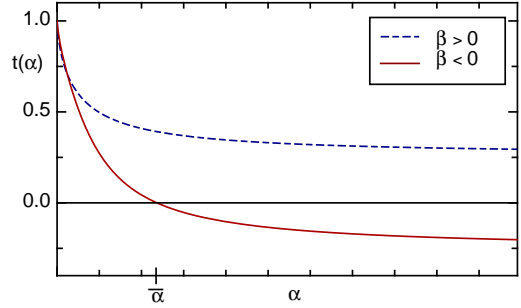}}
\caption{Sketch of the solution $t(\alpha)$  with the pion definition for the critical mass. Depending on the sign of $\beta$ the solution $t$ converges to either a positive or negative value for large $\alpha$. For  $\beta<0$ there exists a finite $\overline{\alpha}$ where $t=0$ and consequently $\owt =\pih$. $t(0)=1$ independently of the sign of $\beta$. }
\label{fig:SolGE}
    \end{center}
    \vspace{-0.3cm}
 \end{figure}

Starting from the gap equation one can derive some general characteristics of $t(\alpha)$. The expected behavior is sketched in fig.\ \ref{fig:SolGE}. For $\alpha\rightarrow 0$ one finds $t \rightarrow 1$, while for $\alpha\rightarrow \infty$ it converges to $t_{-}$, the solution of the equation $t -\beta(1-2t^{2})=0$ which satisfies $|t|<1$. 
The sign of $t_{-}$ is equal to the sign of $\beta$. If $\beta<0$ there exists a value $\overline{\alpha}$ where $t=0$ and $\omega_{\rm WT} = \pi/2$. Increasing the twisted mass beyond $\overline{\alpha}$ results in angles $\omega_{\rm WT} > \pi/2$, and such values are indeed observed in the data \cite{Abdel-Rehim:2005gz}. We emphasize that without the O($\mu a$) terms one finds the bound $\omega_{\rm WT} \le \pi/2$ for all $\alpha$  \cite{Aoki:2004ta}.

The O($\mu a$) contributions are also important for describing another numerical result. 
The parity conservation definition tunes  $m_{0}$ such that $t=0$. If we require  $t=0$ to be a solution of the gap equation we find the condition 
\bea\label{Cond:PConBare}
m_{0} & =&\pm \frac{\beta}{Z_{A}} \mu_{0} + m_{\rm cr}.
\eea
This equation predicts a linear dependence between the bare masses $m_{0}$ and $\mu_{0}$, which is also observed in the data \cite{Abdel-Rehim:2005gz}. 
 %
 \begin{figure}[t]
 \vspace{-0.4cm}
\hspace{1mm}\scalebox{0.6}{\includegraphics{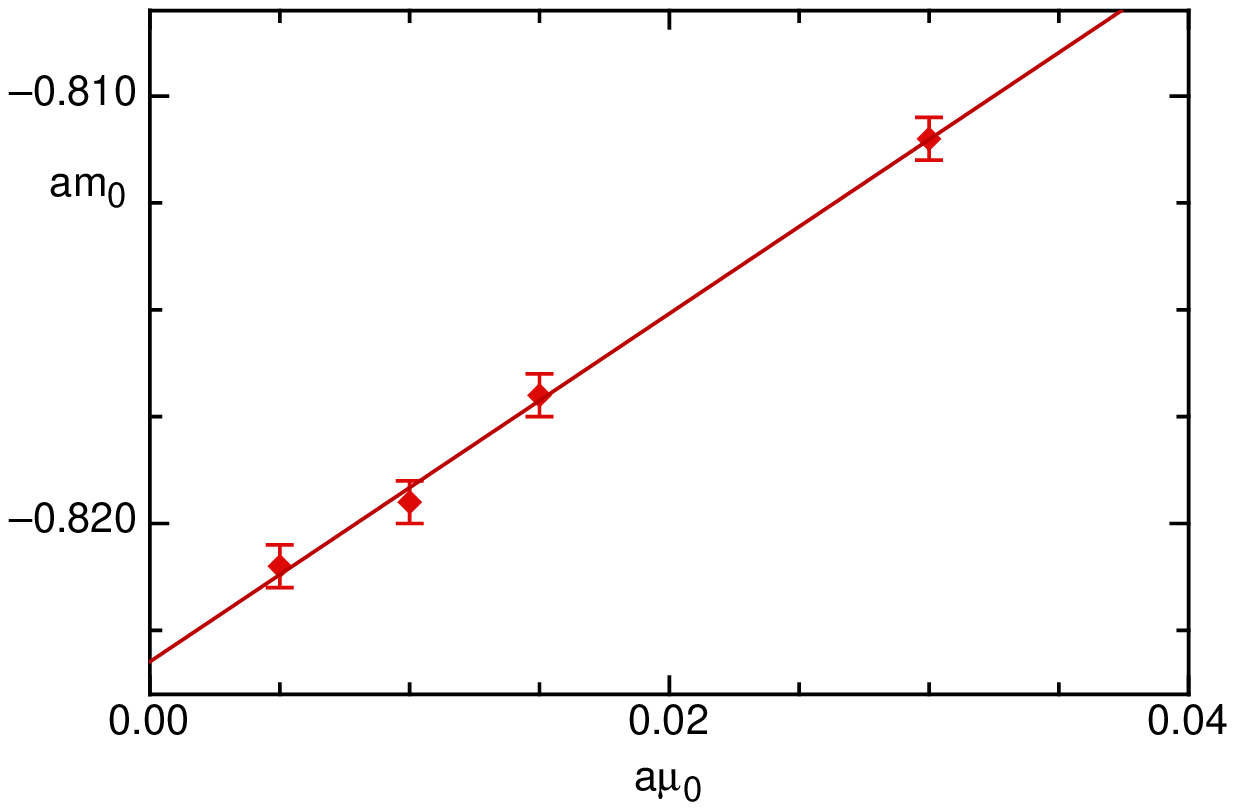}}\hspace{3.5mm}
\scalebox{0.6}{\includegraphics{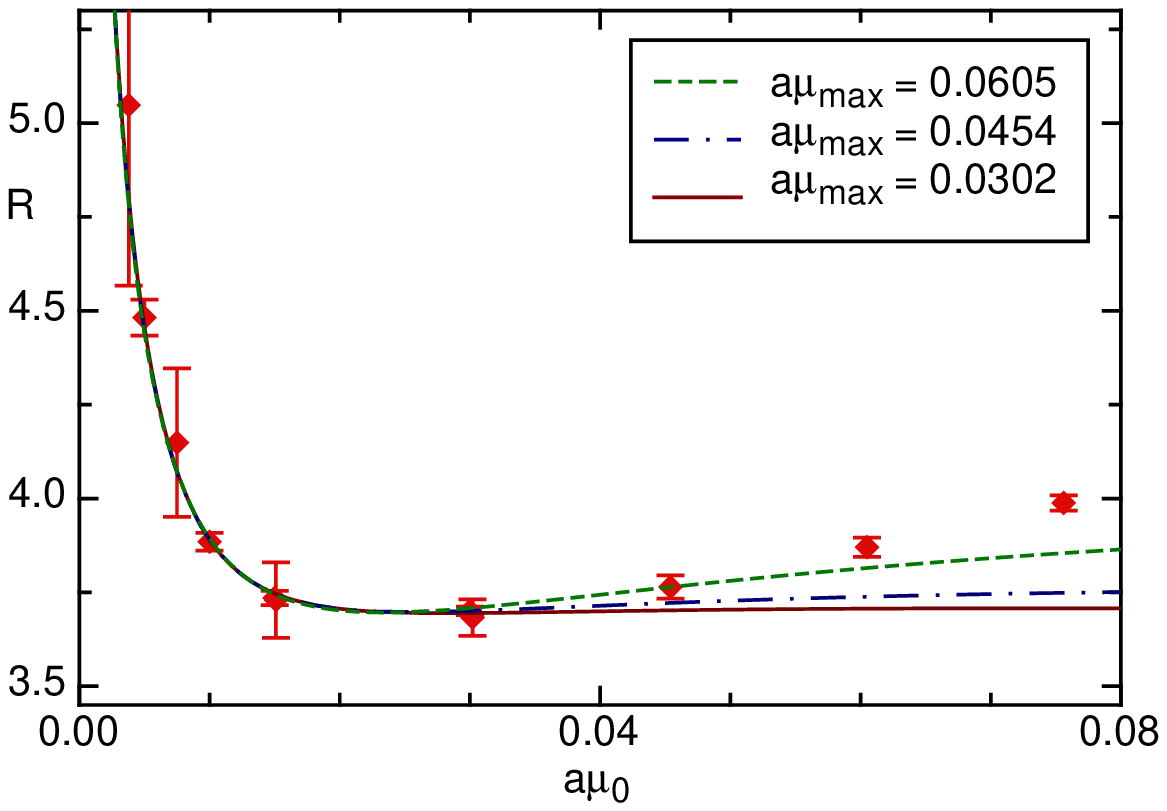}}
\newline
  \scalebox{0.6}{\includegraphics{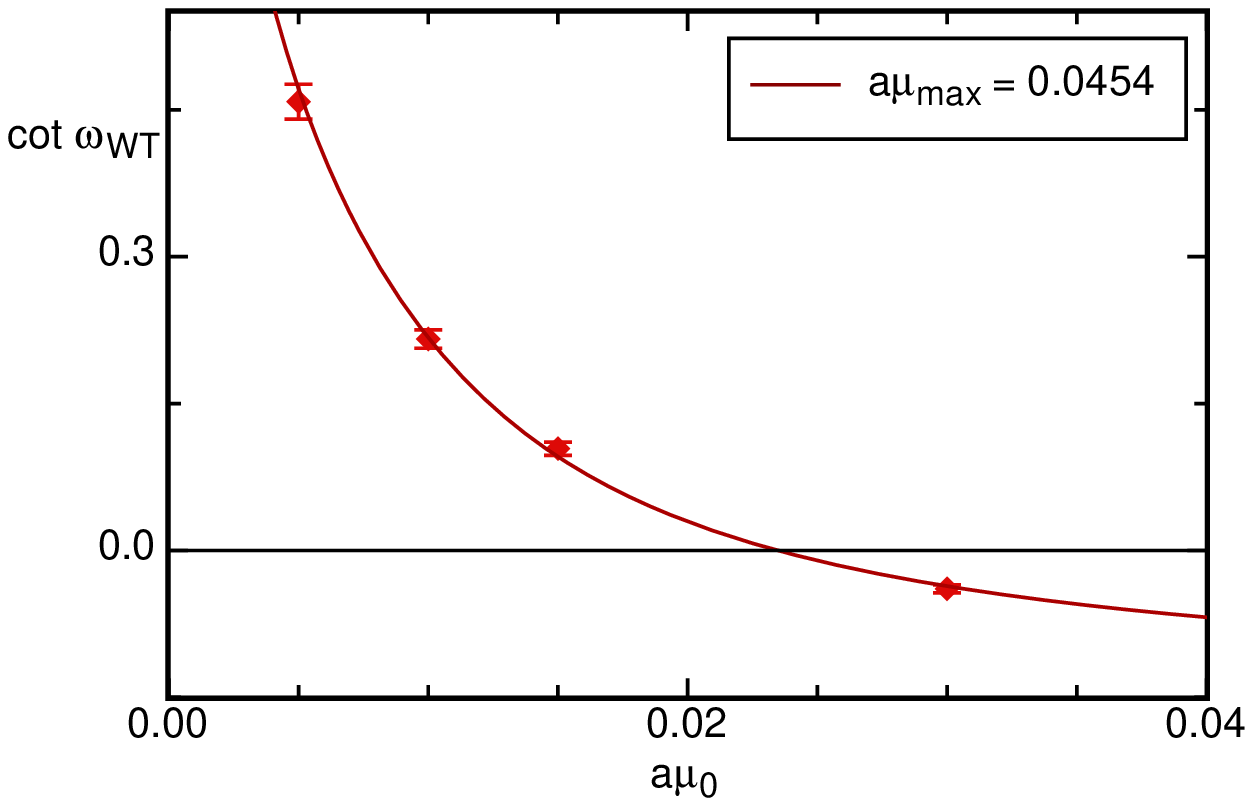}}\hspace{2.5mm}
   \scalebox{0.6}{\includegraphics{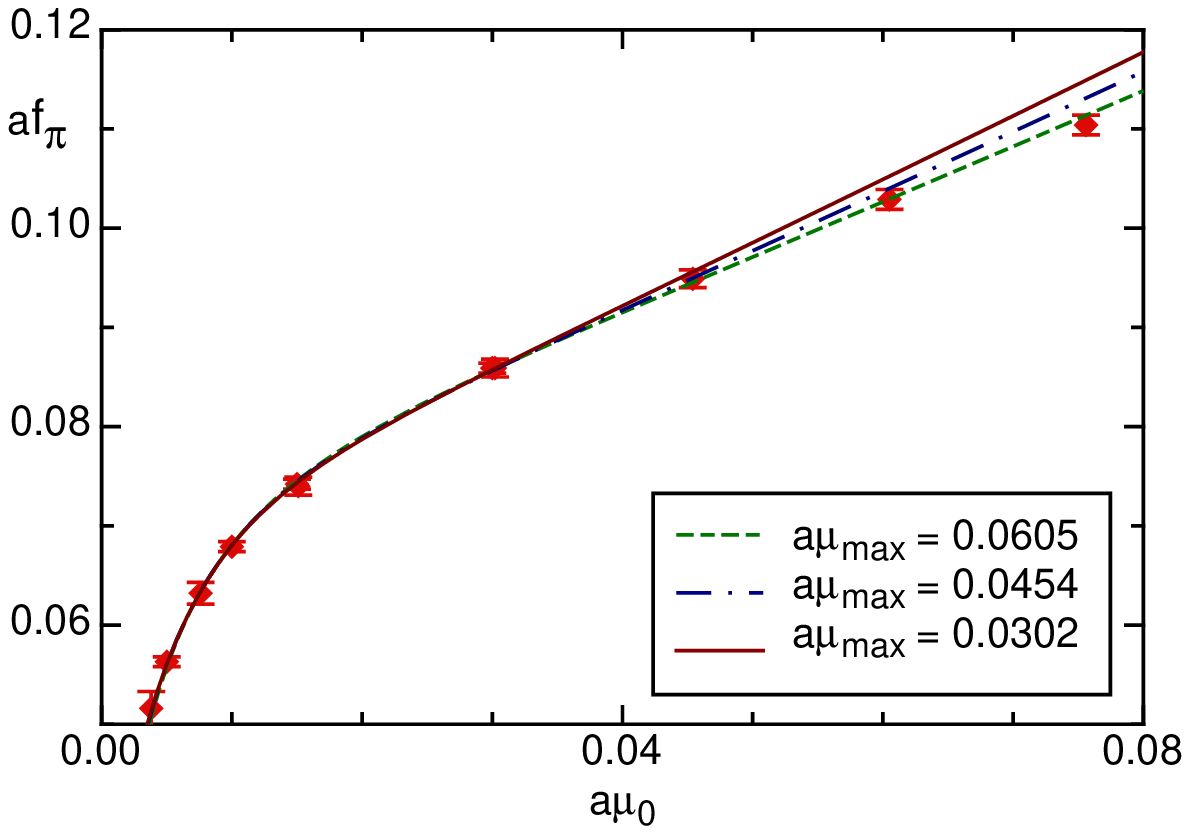}}
  \vspace{-8mm} 
\begin{center}
\caption{Fit results for the $\beta=6.0$ data.  Upper left panel shows $am_{0}$ as a function of $a\mu_{0}$ when tuned according to the parity conservation definition. The remaining panels show  $R= a^{2}m_{\pi}^{2}/a\mu$, $af_{\pi}$ and $\cowt$ for the pion definition. The three lines show the results for different fit ranges when data with  $\mu_{0} \le \mu_{\rm max}$ is included in the fit (only one line is shown for $\cowt$ since all three lines would lie on top of each other).  }
\label{fig:FitResult}
\end{center}
    \vspace{-0.9cm}
\end{figure}
%
%
\section{Analyzing the data}
%
The data in Refs.\  \cite{Bietenholz:2004wv,Abdel-Rehim:2005gz,Jansen:2005gf,Jansen:2005kk} was
generated with the Wilson plaquette action at $\beta =5.85$ ($a\approx 0.123$fm) and $\beta= 6.0$ ($a\approx 0.093$fm). The standard Wilson fermion action with a twisted mass term was used to compute the mesonic quantities. The values for the twisted quark mass cover the range of $m_{\pi}/m_{\rho} \approx 0.3 -0.8$. The untwisted bare quark mass was tuned in two different ways, according to the pion mass and the parity conservation definition for the critical quark mass. 

We performed combined fits  to the data for the ratio $R=a^{2}m_{\pi}^{2}/a\mu$, $af_{\pi}$ and $\cowt$. The fit functions  \pref{Mpi} - \pref{Def:cotw} contain 10 free fit parameters.
A total number of 
34 data points are available for each $a$.
We obtain good fit results with $\chi^{2}/d.o.f \approx 0.3 -1.3$ if we exclude the 
heaviest mass data. Fig.\  \ref{fig:FitResult} displays  some fit results  for $\beta = 6.0$. 
Three fits are shown for $R$ and $af_{\pi}$, depending on the fit range. Obviously, WChPT describes the bending for small quark masses quite well. 

The fit parameters are determined with moderate errors. The parameters in the gap equation, which essentially describe the bending, have statistical errors of O(5\%). A source of larger uncertainty is the choice for the fit range. Depending on data points included in the fit 
the central value for some parameters varies by about 20\%. 
This  suggests that higher order corrections in the chiral expansion are not negligible, even though the results  \pref{Mpi} - \pref{Def:cotw} are able to describe the data quite well. This is not unexpected, since the smallest $\mu_{\rm max}$ in fig.\ \ref{fig:FitResult} corresponds to $m_{\pi}/m_{\rho}\approx 0.65 $, 
which is fairly heavy. 

%
\section{Conclusions}
%
We interpret our preliminary results as follows. 

WChPT is indeed able to describe the bending phenomenon observed in quenched tmLQCD. The origin of the bending is -- in the language of the chiral effective theory -- the contribution of the non-trivial ground state, encoded in the function $t(\mu)$.  The terms of O($\mu a$)  cannot be ignored at the lattice spacings we considered. Once these terms are included WChPT explains a variety of features observed in the data: The angle $\omega_{\rm WT}$ can assume values larger than $\pih$, the untwisted mass depends linearly on the twisted mass when it is tuned according to the parity conservation definition, and the pion mass and decay constant show the strong bending for small quark masses.

Fits to the data provide estimates for particular combinations of low-energy parameters with good statistical precision but sizable systematic errors due to the choice of fit range. The analysis suggests that higher order corrections might be necessary in order to obtain reliable estimates for these parameters with small errors. 
Our final results will be given elsewhere \cite{AokiBar}. 
 \vspace{-0.3cm}

\providecommand{\href}[2]{#2}\begingroup\raggedright\endgroup

\end{document}